# Versatile CMOS modulation-free self-isolating stabilized precision lasers on a chip

David A. S. Heim[1], Kaikai Liu[1], Rahul Chawlani[1], Karl. D. Nelson[2], Daniel J. Blumenthal[1,*]

[1] Department of Electrical and Computer Engineering, University of California Santa Barbara, Santa Barbara, CA, USA
[2] Honeywell Aerospace, Plymouth, MN, USA
* Corresponding author (danb@ucsb.edu)

## ABSTRACT

Ultra-low-noise stabilized lasers are a fundamental tool for precision quantum technologies[1–5], optical clocks[6,7], low-noise microwave and millimeter-wave generation[8–10], and fiber sensing[11]. However, existing systems rely on table-top bulk-optic components, including discrete lasers, reference cavities, isolators, modulators and frequency shifters, limiting their portability, scalability, cost reduction, and manufacturability. While these systems offer flexibility in laser design in order to tailor linewidth, frequency noise, and wavelength to specific applications, fully integrating a stabilized laser onto a chip without sacrificing performance and versatility has remained elusive. Here, we report integration of the precision stabilized laser, consisting of a flexible isolator-free core laser design, and a modulation-free stabilization cavity in the low-loss silicon nitride photonic integration platform. We first demonstrate a stabilized widely tunable self-isolating extended cavity tunable laser (ECTL) monolithically integrated with an on-chip modulation-free coil-loaded Mach-Zehnder interferometer (CL-MZI). This stabilized laser is hybrid integrated with a semiconductor optical amplifier and separate external photodiodes and Pound-Drever-Hall (PDH) electronics. This design yields a fundamental linewidth of 1.7 – 10.5 Hz across a 60 nm tuning range, integral linewidth of 299 – 505 Hz over a 30 nm tuning range, frequency noise reduction of over 5 orders of magnitude, and an Allan-Deviation (ADEV) of $6.5 \times 10^{-13}$ at 0.08 ms. We next highlight the versatility of this approach by demonstrating a monolithic self-isolating stimulated Brillouin scattering (SBS) laser, that provides nonlinear noise suppression of high frequency noise by multiple orders of magnitude, stabilized to an on-chip CL-MZI, with off chip pump laser, photodiodes and PDH electronics. The stabilized SBS laser achieves 4 Hz fundamental linewidth, over an order of magnitude integral linewidth reduction to 74 Hz, and ADEV of $2.8 \times 10^{-13}$ at 5 ms. The stabilization portion of the circuit employs a 4-meter CL-MZI reference cavity that directly generates a PDH error signal for feedback control of the lasers, eliminating the need for an active modulator. These results bring the performance and versatility of table-top stabilized laser systems to a chip for the first time. The ultra-low loss and wavelength transparency of silicon nitride enables ready translation of this approach to operation from the visible through shortwave-IR with further noise and integral linewidth reduction and improved stability through increased coil-length[12] and resonator quality factor (Q)[13,14]. These monolithic laser systems on chip can be readily integrated with semiconductor optical amplifiers, pump lasers, photodiodes, and field programmable gate array electronic die, to provide hybrid and heterogenous, CMOS foundry compatible stabilized laser solutions, unlocking the path to scalable, low-cost, and manufacturable precision lasers for portable quantum, sensing, and communications applications.

# INTRODUCTION

Stabilized narrow linewidth lasers enable a wide range of applications, including quantum sensing and computing[1–4,15], optical atomic clocks[6,7], ultra-low-noise microwave generation[8,10], gravitational sensing[5,16], new particle discovery[17,18], precision fiber synchronization and coherent communications[11,19], and fiber-optic sensing[20]. Lab-scale stabilized lasers are realized by Pound-Drever-Hall (PDH) locking a low noise table-top laser to a bulk-optic ultra-stable optical cavity using feedback electronics and other bulk optic components including isolators, modulators, and frequency shifters such as acousto-optic modulators (AOMs). Such systems provide a large degree of flexibility to the experimentalist through the ability to choose different laser designs based on matching their optical parameters to the particular experiment, making the stabilized laser a ubiquitous tool for precision and quantum sciences. Progress has been made to reduce the size of these systems primarily through miniaturization of the bulk-optic reference cavity[21,22]. Integration of the ultra-low-noise laser, optical reference cavity, isolators, and other stabilization components onto a single platform would reduce size, weight, and power consumption, while improving portability, cost, and manufacturability. Although individual elements of precision stabilized lasers have been integrated, full implementation of a PDH-stabilized laser on a single chip, and particularly in a way that retains the design flexibility of table-top systems, has remained elusive.

The ultra-low loss silicon nitride CMOS foundry compatible ($Si_3N_4$) platform[23] is a leading candidate for visible (VIS) to shortwave IR (SWIR) integrated photonics[24–28] and quantum systems on chip[29]. Ultra-low fundamental linewidth (FLW) and low noise lasers include self-injection locking (SIL)[30–32], extended cavity tunable (ECTL)[33–35], and stimulated Brillouin scattering (SBS)[14,36,37] laser designs[23,38]. Integrated high quality factor (Q) waveguide reference cavities based on coil[25,39] and spiral resonator[40] geometries provide large extinction ratios (ER) and optical mode volumes, enabling suppression of the thermorefractive noise (TRN) floor, narrowing of the integral linewidth (ILW) and improved fractional frequency stability. A variety of on-chip modulation approaches have been demonstrated in silicon nitride, such as integrated stress-optic actuators in lead zirconate titanate (PZT)[41,42] and aluminum nitride (AlN)[43], and electro-optic modulators in lithium niobate[44,45]. Integrated AOM frequency shifters have also been demonstrated[46,47] yet remain power-hungry and challenging to integrate. Modulation-free stabilization offers a simpler alternative and has been explored using both bulk-optic and integrated cavities[48–51]. Yet, despite progress using separate integrated photonic chips such as Brillouin and ECTL lasers PDH-locked to coil resonator chips[24,52], these systems continue to depend on external modulators, frequency shifters, optical isolators, and other optics and photonics, preventing single chip solutions.

Here, we report the first demonstration of a stabilized precision laser on a single chip, replacing what normally requires a table top of assorted laser and photonic components (see Fig 1). The chips are fabricated in a 200 mm silicon nitride CMOS foundry compatible photonics integration process. The monolithically integrated chips contain a self-isolating precision laser, a modulation-free coil-loaded interferometer reference cavity, and other passive photonic circuitry required for stabilization and self-isolation. Two classes of PDH-stabilized lasers are realized using these monolithic chips in combination with hybrid integrated semiconductor gain and fiber connected laser pump, photodiodes, and PDH locking circuitry (Fig. 1), that can be further integrated with the monolithic chip using hybrid and heterogeneous integration and packaging approaches. First, we demonstrate a laser design that has wide tunability using a self-isolating extended cavity tunable laser (ECTL) stabilized to an on-chip modulation-free coil-loaded Mach-Zehnder interferometer (CL-MZI). The stabilized ECTL is tunable over nearly 60 nm, achieving low FLWs of between 1.65 – 10.48 Hz and stabilized ILWs of 299 – 505 Hz measured over a 30 nm range and an Allan deviation (ADEV) of $6.5 \times 10^{-13}$ at 0.08 ms. Next, we demonstrate versatility with a second core laser design that provides nonlinear high frequency noise suppression using a self-isolating stimulated Brillouin scattering (SBS) laser stabilized to an on-chip CL-MZI. The stabilized SBS laser chip delivers 9 mW output power with a low 4 Hz FLW, a 20X reduction in ILW to 74.2 Hz, and an ADEV of $2.8 \times 10^{-13}$ at 5 ms. The photonic stabilization circuit portion of both chips consists of a 4-meter CL-MZI that provides modulation-free frequency-discrimination, directly generates the PDH locking error optical signal, and in combination with off chip photodiodes and PDH electronic circuitry enables laser frequency noise reduction by six orders of magnitude. Together, these results constitute the first demonstrations of PDH locked frequency-stabilized lasers on a single photonic integrated chip, establishing a scalable, compact, and robust solution for deployable precision measurement and quantum technologies.

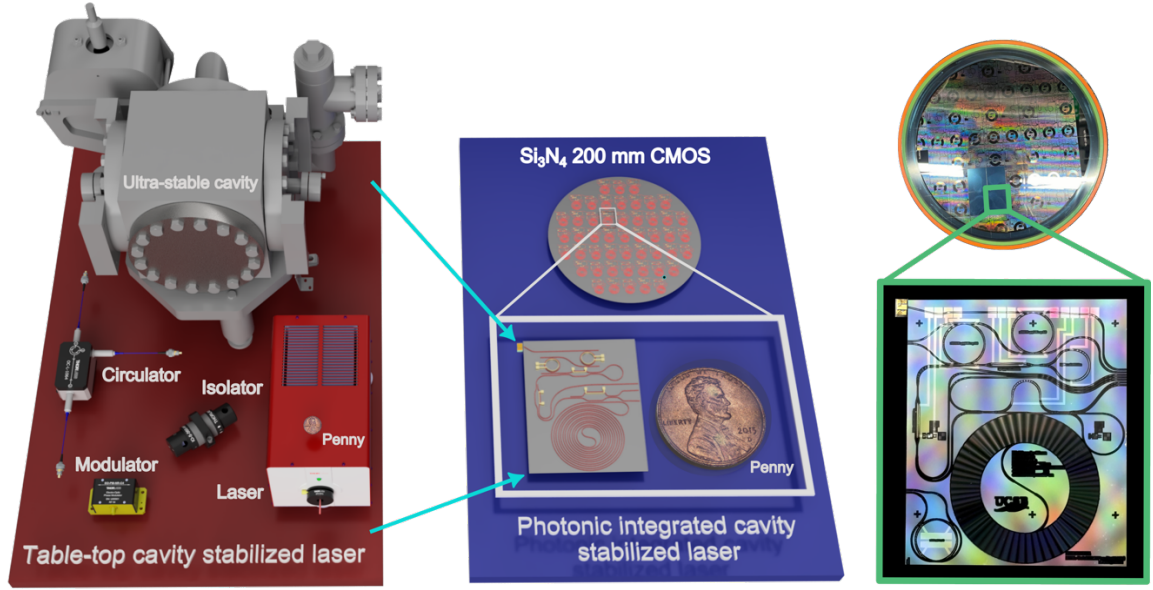

**Fig. 1 | Photonic integration of precision cavity-stabilized laser systems.** Bulk optic components and ultra-stable reference cavities used in table-top stabilized lasers are replaced by a chip-scale implementation fabricated in a 200-mm CMOS photonic platform, reducing size, weight, cost, and complexity. On the right are photographs of the fabricated wafer and one of the stabilized extended cavity tunable laser (ECTL) devices, including the hybrid-integrated gain chip.

# RESULTS

### Stabilized laser architecture

Our integration approach, based on the ultra-low-loss silicon nitride ($Si_3N_4$) platform, supports high performance precision laser and stabilization photonics using a set of functional building blocks including ultra-low loss waveguides, high-Q resonators that enable long photon lifetimes and enhance nonlinear effects, double-bus ring resonators for narrowband filtering, Sagnac loop mirrors for tunable broadband reflection, tunable MZIs for routing, passive splitters and combiners, and thermo-optic phase actuators. The versatile monolithic design is shown in Fig. 2a. The lasers are realized using a narrow linewidth silicon nitride laser toolkit (middle Fig. 2a) and are stabilized using an on-chip coil-loaded MZI. Based on the design of the silicon nitride lasers, the monolithic chips are connected to off-chip hybrid integrated reflective semiconductor optical amplifier (RSOA) for the ECTL or fiber coupled semiconductor laser pump for the SBS laser. Both chip designs produce the optical PDH error signal at the output which are connected to off chip balanced differential photodetectors (right Fig. 2a) and processed by off chip locking electronics. The semiconductor gain and laser and photodiodes can be readily integrated via hybrid or chiplet heterogeneous integration using well developed flip-chip bonding, pick and place, or micro transfer print (MTP) techniques.

The coil-loaded MZI consists of a meter-scale coil resonator whose large mode volume reduces the thermorefractive noise (TRN) floor[12,53] and narrows the ILW, while the long path length yields a closely spaced free spectral range (FSR) that enables laser locking on a fine frequency grid across a wide tuning range. The phase of the delay-line waveguide is thermo-optically tunable to maximize the slope of the error signal. Balanced detection of the coil-loaded MZI, using fiber coupled detectors, of the coil-loaded MZI yields a high-discrimination Fano resonance error signal (arising from interference between the resonator and the MZI path difference) for modulation-free stabilization[48] via direct frequency feedback to the laser.

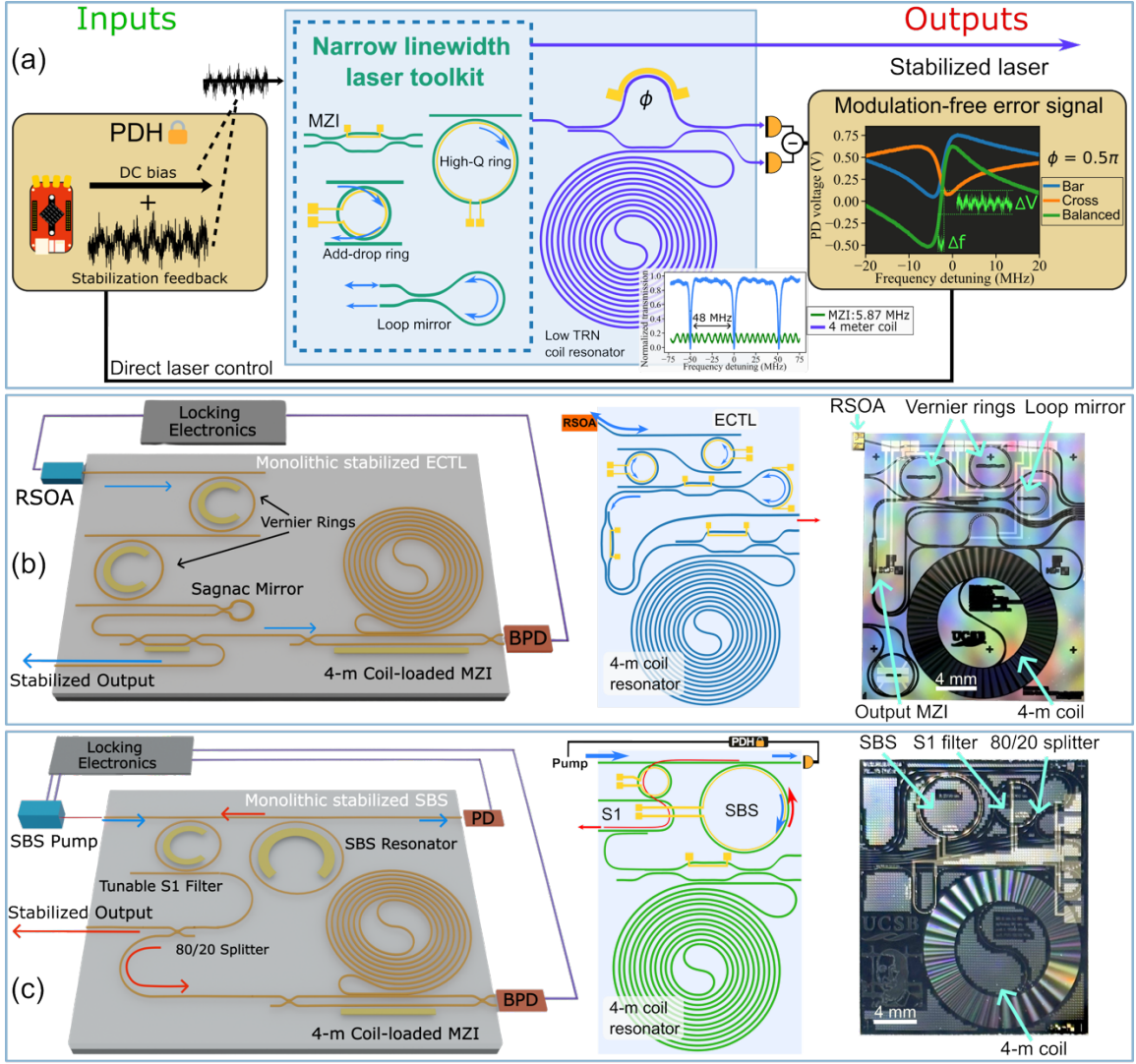

**Fig. 2 | Integrated stabilized laser architecture. a** An integrated narrow linewidth laser built from silicon nitride ($Si_3N_4$) components is stabilized by probing a 4-meter coil-loaded Mach Zehnder Interferometer (MZI) integrated on the same photonic chip. The large mode volume resonator serves as a low thermorefractive noise (TRN) reference cavity. Balanced detection of the two outputs produces a high-discrimination, asymmetric error signal for laser locking via direct feedback to the laser frequency. A micro-heater on the delay-line arm of the MZI adjusts the phase delay (ϕ) to optimize the error signal slope. The transmission spectrum of the 4-meter coil resonator with 48 MHz free spectral range (FSR) is shown, creating a fine wavelength grid for locking the extended cavity tunable laser (ECTL) over a wide tuning range. The stabilized ECTL chip has an electrical input of a DC bias current plus the stabilization feedback signal and it outputs three optical signals: two from the MZI, that when balanced detected generate the modulation-free error signal, and the stabilized laser output. **b** Experimental layout, device design, and photograph of the monolithic stabilized ECTL chip with the hybrid-integrated reflective semiconductor optical amplifier (RSOA) coupled in the top left corner. **c** Experimental layout, device design, and photograph of the monolithic stabilized Brillouin laser chip with the pump laser connected via optical fiber.

## Monolithic stabilized ECTL

Using this architecture, we integrate two different stabilized laser designs. The first chip is a self-isolating widely tunable extended cavity tunable laser (ECTL) that is monolithically integrated with a 4-meter coil-loaded MZI, fabricated in a planar 80 nm thick $Si_3N_4$ CMOS process (see Methods on monolithic $Si_3N_4$ photonic circuit fabrication). The ECTL employs dual Vernier-style intracavity ring resonators together with thermally tunable phase control and a tunable Sagnac loop mirror. The intracavity rings, with loaded Q factors of 1.65 million at 1550 nm, increase the effective cavity length and photon lifetimes thereby reducing the fundamental linewidth and provide self- isolation of over -30 dB[52], relative to that of a commercial III-V DFB laser, enabling integration of the laser and stabilization circuit without an optical isolator. Thermo-optic tuning of the rings enables wide tuning across nearly 60 nm (Fig. 3a), and we measure SMSRs of up to 65 dB. The ECTL exhibits a lasing threshold of 70 mA and delivers 1 mW of fiber-coupled off-chip output power measured at 1540 nm with an RSOA current of 220 mA. See Supplementary Information Sec. 3 for more details. An integrated tunable MZI at the ECTL output selectively routes a portion of the laser output to the coil-loaded MZI stabilization circuit. The chip is then hybrid packaged with a butt-coupled reflective semiconductor optical amplifier (RSOA), fiber-coupled differential photodetectors, and a non-integrated electronic current driver and PDH-servo controller (Vescent D2-125). The stabilized ECTL experimental layout, with hybrid integrated RSOA, off-chip photo-detectors, and off-chip PDH stabilization electronics, device design, and photograph of fabricated chip are shown in Fig. 2b, left to right, respectively.

## Monolithic stabilized SBS

The second chip is a narrow linewidth stimulated Brillouin scattering (SBS) laser[54] monolithically integrated with a 4-meter coil-loaded MZI, fabricated in the same 80 nm thick planar $Si_3N_4$ CMOS process. SBS lasers have a unique fundamental linewidth narrowing property that leverages photon-phonon interactions[55] and are capable of producing stimulated emission with sub-hertz-level fundamental linewidth[56,57] and low frequency noise at high frequency offsets. The design is self-isolating by combining the non-reciprocity of Brillouin scattering with an integrated tunable first-order Stokes (S1) add-drop ring filter that separates the pump and SBS output, providing 25 dB of self-isolation and eliminating the need for a fiber circulator to extract the output Stokes signal. The SBS resonator has a loaded 16.2 million and intrinsic 39.0 million Q at 1550 nm with a linewidth of 12 MHz, corresponding to a waveguide propagation loss of 0.8 dB/m. The SBS laser is optically pumped by an external cavity diode laser (ECDL) that is frequency-locked to the SBS resonance. The threshold is measured to be 6 mW and the laser outputs 9 mW with a pump power of 25 mW. The SBS output is measured on an OSA, shown in Fig. 3a. The on-chip filter ring is thermally tuned to align with the S1 tone and results in 29 dB of side mode suppression (SMSR) due to the residual pump power. See Supplementary Information Sec. 2 for more details. An integrated 80/20 evanescent splitter selectively routes 20% of the laser output to the coil-loaded MZI stabilization circuit. The pump laser, differential photodetectors, and stabilized laser output are all connected via optical fiber, and a non-integrated electronic current driver and PDH-servo controller is used to control the pump laser and execute the laser lock. The stabilized SBS laser experimental layout, device design, and photograph of fabricated chip are shown in Fig. 2c, left to right, respectively.

## Frequency noise and modulation-free laser stabilization

The experimental setup for the measuring the frequency noise (FN) is outlined in Fig. 3a. We characterize the frequency noise, linewidth and Allan deviation (ADEV) of the stabilized lasers by using both an optical frequency discriminator (OFD) in a self-delayed homodyne configuration and a stable reference laser (SRL) for heterodyne beatnote detection. Without the need for isolation or modulation, the two outputs of the coil-loaded MZI are balanced detected to generate the PDH-locking error signal that is fed back to stabilize the laser frequency. The 4-meter coil resonator, with a loaded Q of 28 million and a 48 MHz FSR, serves as a low noise frequency discriminator, and we measure balanced detected error signal slopes of between 0.14 and 0.53 MHz/V across the operating wavelengths (see Supplemental Information Sec. 4). When locked, the lasers inherit the frequency noise characteristics of their respective 4-m coil-loaded MZI frequency discriminators for offset frequencies within the PDH locking bandwidth. The two independent techniques for measuring frequency noise utilized here are described in Methods and in refs[39,52]. The SBS laser stabilization employs pump laser power modulation for fast frequency modulation and an on-chip microheater for slow and auxiliary frequency modulation feedback, resulting in a lock loop bandwidth of 0.8 MHz. The ECTL laser stabilization feeds back to the RSOA current for high-speed frequency modulation and achieves a lock loop bandwidth of 0.1 MHz. Further details of the PDH locking electronics are given in the Method Section.

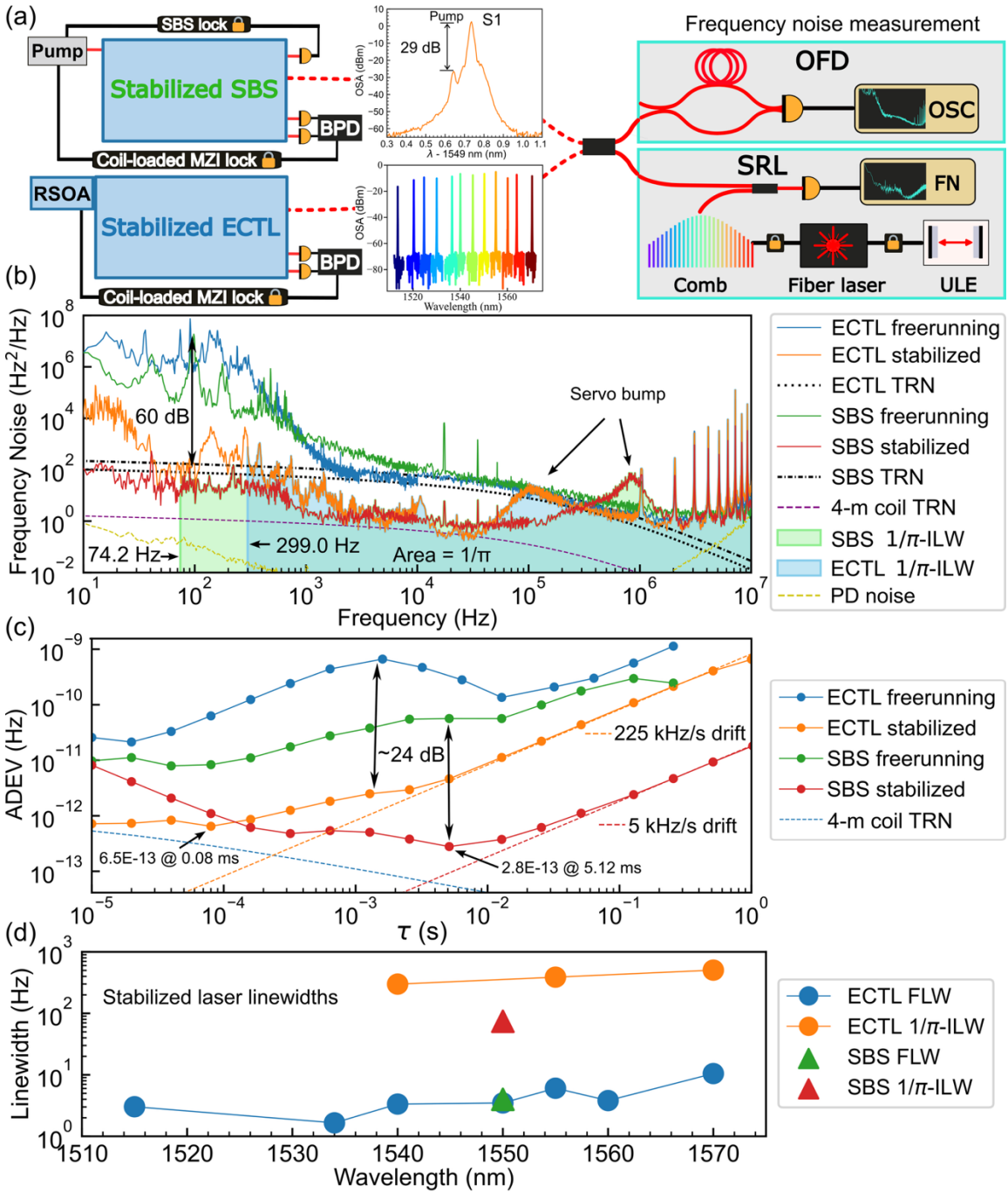

**Fig. 3 | Stabilized laser frequency noise measurements. a** The stabilized ECTL and SBS lasers are Pound-Drever-Hall (PDH) locked by measuring the optical outputs of their respective on-chip 4-m coil-loaded MZI's using a balanced photodetector (BPD). The SBS laser output (S1) is measured on an optical spectrum analyzer (OSA) showing a side-mode suppression ratio (SMSR) of 29 dB due to the suppression of residual pump power with the on-chip tunable filter ring. The ECTL output is measured on an OSA for different ring heater currents across the nearly 60 nm tuning range. The frequency noise (FN) of the lasers is characterized using two independent techniques: high frequency noise is measured using a 1.026 MHz unbalanced fiber MZI as an optical frequency discriminator (OFD) and low frequency noise measured down to 1 Hz is characterized using a heterodyne beatnote with a stable reference laser (SRL) locked to an optical frequency comb. **b** FN of the free running and stabilized lasers, showing 5.15 Hz fundamental linewidth (FLW) and 299.0 Hz integral linewidth (ILW) for the stabilized-ECTL operating at 1540 nm, and 3.99 Hz FLW and 74.2 Hz ILW for the stabilized-SBS laser. They both show ~6-orders of magnitude of FN suppression near 100 Hz frequency offset, and approach the thermorefractive noise (TRN) floor of the 4-meter coil resonator between 1 – 100 kHz frequency offset. **c** Allan deviation (ADEV) of the free running (ECTL: blue, SBS: green) and stabilized lasers (ECTL: orange, SBS: red), show ~24 dB reduction in fractional frequency stability for each, and a minimum at $6.5 \times 10^{-13}$ at 0.08 ms and $2.8 \times 10^{-13}$ at 5.12 ms for the stabilized ECTL and stabilized SBS, respectively. **d** Summary of the measured fundamental and integral linewidths of the stabilized lasers as measured across their tuning ranges.

The frequency noise spectra of the stabilized ECTL and SBS lasers are plotted in Fig. 3b. The free-running (blue) and cavity-stabilized ECTL (orange), operating at 1540 nm, and the free-running (green) and cavity-stabilized SBS (red) each show a FN reduction of approximately 6-orders of magnitude near 100 Hz frequency offset, and approach the TRN limit of the 4-meter coil resonator cavity between 1 and 500 kHz offset, before sloping upwards due to the servo bumps at 0.1 and 0.8 MHz, respectively, corresponding to the PDH locking bandwidth. The SBS fundamental linewidth (FLW) is measured to be 3.99 Hz (green-shaded region) a reduction of > 20x from free-running. The ECTL FLW ranges from 1.65 to 10.48 Hz measured across a wavelength range of 1515 – 1570 nm. We measure the $1/\pi$ integral linewidth (ILW), calculated using the reverse integration method, of the stabilized SBS to be 74.2 Hz and the ILW of the stabilized-ECTL to be 299.0 Hz as shown by the blue-shaded region, representing a reduction of >3X from free-running. We also measure the stabilized-ECTL ILW at three points across the tuning range to be 299 – 505 Hz measured from 1540 – 1570 nm. The 48 MHz FSR of the 4-m coil reference cavity readily enables locking at any point across the ECTL 60 nm tuning range. The ADEV and drift of the stabilized-ECTL and stabilized-SBS lasers are plotted in Fig. 3c, and measured down to 1 Hz showing ADEV improvement of ~ 24 dB. The stabilized-ECTL ADEV is $6.5 \times 10^{-13}$ at 0.08 ms with 225 kHz/s drift and the stabilized-SBS has an ADEV of $2.8 \times 10^{-13}$ at 5.12 ms with 5 kHz/s drift. A summary of the fundamental and integral linewidths of the stabilized lasers is shown in Fig. 3d.

## DISCUSSION

This work demonstrates, for the first time, a reference-cavity PDH stabilized laser in which the laser, the reference cavity, and the stabilization photonics are monolithically integrated on the same photonic chip, demonstrating the ability to replace table top stabilized lasers with chip-scale integrated photonics. The co-integration, in a 200 mm CMOS platform, is achieved with self-isolating lasers and a modulation-free optical stabilization circuit, enabling a high performance and robust approach to fully integrated precision laser systems. Additionally, we demonstrate modularity and design diversity that has been the hallmark of table top systems by demonstrating stabilized ECTL and stabilized Brillouin laser designs that highlight the unique characteristics of each laser, namely wide tunability and nonlinear noise reduction, respectively. We utilize the same modulation-free 4-meter coil-loaded MZI locking scheme to two the fundamentally different integrated laser types, achieving hertz-level fundamental linewidths and sub-kHz integral linewidths.

Demonstration of both laser types can be stabilized using an identical on-chip reference cavity highlights the generality of the technique and suggests that a common stabilization architecture can support a broad range of integrated laser designs, rather than being tailored to a single gain or feedback mechanism.

The low propagation loss of the silicon nitride platform, measured here to be < 1 dB/m at 1550 nm, enables high-Q resonators that allow long photon lifetimes, efficient nonlinear scattering, sharp frequency discrimination, high optical power handling, lowered TRN noise floors, and isolator-free operation. The result is a single stabilized laser chip that emits spectrally pure light with ILWs of 74 Hz and 299 Hz and fractional frequency stability of $2.8 \times 10^{-13}$ at 5.12 ms and $6.5 \times 10^{-13}$ at 0.08 ms, respectively, for the stabilized-SBS and stabilized-ECTL chips. The better performance of the stabilized-SBS device in this instance is likely due to the packaging used which provided better vibration and acoustic isolation as compared to the stabilized-ECTL which will improve with packaging of the edge-coupled RSOA such as photonic wire bonding (PWB)[58].

While these chip designs operate in the shortwave IR (SWIR), the silicon nitride platform has been shown to be low loss and can realize these elements from the visible to SWIR (410 nm – 2350 nm)[24,31,36,59,60] hence this work is readily scalable to the visible and near IR (NIR) spectrum with the appropriate semiconductor gain media and photodetectors.

Also, given the flexibility of this platform and architecture, other configurations are possible, and are a focus of future work. For example, the ECTL could serve as a pump for the SBS laser, enabling strong suppression of high-frequency noise through the Brillouin process while the on-chip coil resonator reduces low-frequency noise through reference-cavity locking. This layered noise suppression – combining nonlinear linewidth narrowing with cavity stabilization – demonstrates how integrated photonics enables composite, phase stable laser architectures that would be difficult to realize in bulk or fiber-based platforms. The performance of the stabilized lasers is closely tied to the properties of the integrated reference cavity. Coil resonators fabricated in ultra-low-loss $Si_3N_4$ provide large optical mode volumes and correspondingly low TRN floors (TRN scales inversely with optical mode volume), with on-chip cavity lengths of up to 17 meters, while maintaining 250 million Q, have already been demonstrated[12]. Extending the present approach to longer coils offers a direct route to further reductions in low frequency noise and integral linewidths, and greater stability, without altering the laser or stabilization circuitry.

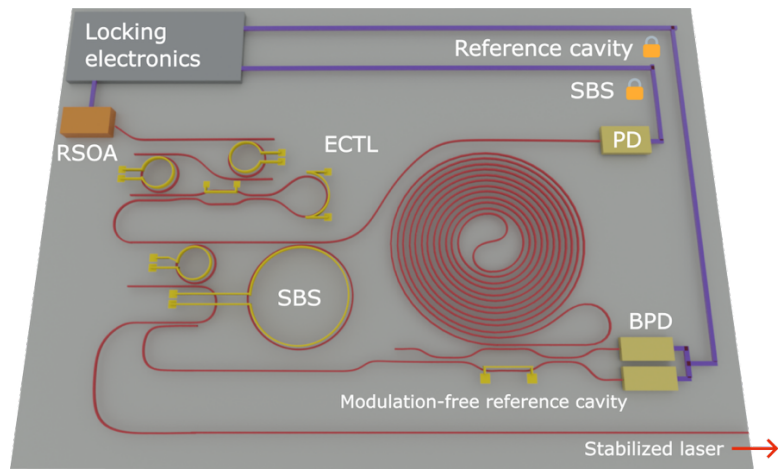

**Fig. 4 | ECTL-pumped stabilized SBS laser.** Future designs can utilize the ECTL as an integrated pump laser, the SBS laser as a second stage noise filter to reduce high frequency noise, and then the same modulation-free coil-loaded MZI lock to bring down low frequency noise. Photodetectors, CMOS electronics, and gain medium are all available for integration with the low loss silicon nitride platform.

Beyond linewidth and stability, the architecture is compatible with additional functionality needed for deployable systems. The lasers and integrated reference cavity can be readily combined with piezoelectric[41] or electro-optic actuators[28,61] to provide fast and wide-range cavity tuning with low cross talk, enabling agile frequency control[62] and compensation for environmental drift[63]. Additional on-chip or hybrid-integrated gain stages can also be incorporated to increase output power while preserving the stabilized spectral properties[64]. Photodiodes can also be incorporated on-chip[65] for the PDH-locking balanced detection, as well as CMOS electronics for the locking servo[66,67], enabling a full heterogeneously-integrated and CMOS-compatible solution. More broadly, the demonstrated low noise lasers and reference cavities on a single photonic chip opens a pathway toward chip-scale precision lasers suitable for field-deployable applications. By eliminating bulk cavities, free-space optics, and external modulation hardware, this work bridges the gap between laboratory-grade ultra-stable lasers and integrated photonic sources. Such systems are poised to impact precision sensing, coherent communications, fiber sensing, and quantum technologies, where compactness, robustness, and scalability are critical.

# METHODS

**Monolithic silicon nitride photonic circuit fabrication:** The silicon nitride ($Si_3N_4$) photonic chips are fabricated in a CMOS foundry. First, a 15-μm-thick thermal oxide lower cladding layer is grown on a 200-mm diameter silicon substrate wafer. A 75-nm-thick $Si_3N_4$ film is deposited on the thermal oxide using low-pressure chemical vapor deposition (LPCVD), followed by a standard deep ultraviolet (DUV) photoresist spinning, DUV stepper patterning and dry etching in an inductively coupled plasma etcher using $CHF_3/CF_4/O_2$ chemistry. Following the etch, a standard Radio Corporation of America (RCA) cleaning process is applied. An additional $Si_3N_4$ thin layer is deposited followed by a 30-minute and 1100 °C anneal in an oxygen atmosphere. Lastly, a 6-μm-thick silicon dioxide upper cladding layer is deposited using plasma-enhanced chemical vapor deposition (PECVD) with tetraethoxysilane (TEOS) as a precursor, followed by a final two-step anneal at 1050 °C for 7 hours and 1150 °C for 2 hours.

**Circulator-free and isolator-free stabilized SBS laser:** The waveguide design of 6 μm wide and 80 nm thick silicon nitride waveguides strikes a good balance between the propagation loss and waveguide confinement that results in a bending radius below 1 mm and enables putting the large-mode-volume 4-meter-coil resonator, the SBS resonator, the add-drop filter resonator and the 80/20 splitter on the same chip with a footprint of 21 mm by 26 mm. The lower thermal oxide cladding is 15 μm thick, and the upper cladding oxide is 6 μm.

To remove the need for a fiber optic circulator to drop the SBS laser output from the reflection port, an add-drop filter resonator that uses the same bus waveguide as the SBS resonator can selectively drop the S1 to the drop-port bus waveguide, which also filters out any back-reflected pump (see Supplementary Information Sec. 2). The SBS ring resonator has a radius of 2.770 mm and a bus-ring coupling gap of 2.0 μm so that it is slightly over-coupled at 1550 nm for better pump-Stokes power conversion efficiency, while the filter ring resonator has a radius of 2.0 mm and a bus-ring coupling gap of 0.5 μm so that the add-drop filter ring resonator is heavily over-coupled and a very minimal add-drop insertion loss. Both the filter and SBS ring resonators have a micro-heater on top of them to tune the optical resonances thermally.

To operate the SBS laser with the on-chip add-drop filter, the filter ring resonator is thermally tuned to align with the SBS resonance at which the S1 emission is located. Then, the SBS laser is optically pumped by an external cavity diode laser that is frequency-locked to the SBS resonance. The SBS threshold is measured to be 6 mW and the SBS laser outputs 9 mW with a pump power of 25 mW at the S1 clamping point. We compare the performance of the on-chip add-drop filter as compared to a fiber circulator, and also directly measure the SBS laser's resilience to optical feedback: for results see Supplementary Information Sec. 2.

**Stabilized ECTL laser:** The ECTL laser design is also in 80 nm thick silicon nitride with the intracavity rings having radii of 1998.36 and 2002.58 μm, waveguide widths of 2.8 μm, and bus-ring gaps of 2.75 μm. The waveguide tapers out to 18 μm wide at the facet for edge coupling to the RSOA (Thorlabs SAF 1128C) and utilizes the fundamental TE mode. Adjusting the micro-heaters on the intracavity rings tunes the laser output with a measured average tuning efficiency of 65 nm/W (see Supplementary Information Sec. 3). After the ECTL the waveguide width increases to 6 μm so the tunable-MZI and 4-meter coil-loaded MZI have the same waveguide dimensions as in the stabilized SBS chip, and all fit in the same 21 x 26 mm footprint.

**Frequency noise measurements:** Both the free running and stabilized laser outputs are measured by two independent methods. A 200-meter-delayline fiber MZI with an FSR of 1.03 MHz is used as an optical frequency discriminator (OFD) and we measure the self-delayed homodyne FN signal on a high-speed balanced photodetector. Excess fiber noise in the MZI OFD, particularly below 1 kHz frequency offset, requires introducing a second FN measurement method. For low offset frequency noise the laser output is mixed with and measured against a stable reference laser (SRL) that is stabilized to an ultra-low-expansion and high-finesse Fabry–Pérot cavity and has a 1-hertz-level integral linewidth and ~0.1 Hz/s frequency drift[52]. The heterodyne beatnote signal is measured on a high precision frequency counter. To measure the widely tunable ECTL, we lock a Vescent self-referenced fiber frequency comb with a 100 MHz $f_{rep}$ to the SRL to extend the stability of the SRL system to many wavelengths and enable accurate close-to-carrier FN measurements across the ECTL spectrum.

**Locking electronics:** Laser stabilization is implemented using a Thorlabs fixed-gain balanced photodetector (BPD), whose RF output is connected directly to a Vescent D2-125 laser servo. The error signal is derived from the Fano resonance of the coil-loaded MZI, enabling modulation-free locking without the need for dither or demodulation. The D2-125 servo output drives a Vescent D2-105 current source for frequency actuation. No additional RF conditioning elements are used in either lock loop. For the SBS laser, fast frequency feedback is provided via pump power modulation, which steers the SBS laser optical frequency. A thermo-optic microheater on top of the SBS ring resonator is used as a slow and auxiliary feedback mechanism allowing the pump power to remain near its operating point and reduce SBS laser intensity fluctuations. The upper bound of the stabilization bandwidth is set by the 180° phase lag frequency of the pump-modulated SBS frequency response, measured to be 0.8 MHz. We measure the SBS laser frequency response directly using a fiber OFD and vector network analyzer (VNA); full details of this characterization are provided in Supplementary Information Sec. 6. For the ECTL, frequency feedback is applied directly to the RSOA injection current, achieving a loop bandwidth of 0.1 MHz.

# DATA AVAILABILITY

The data that support the plots within this paper and other findings of this study are available from the corresponding author on reasonable request.

## ACKNOWLEDGEMENTS

This work is supported by funding from DARPA MTO GRYPHON program under award HR0011-22-2-0008, the Army Research Office (ARO) AMP program under award W911NF2310179, and the NSF under QLCI QsENSE award number 2016244. The authors thank Dr. Andrei Isichenko for productive discussions and feedback on locking servos and feedback electronics.

## DISCLAIMER

The views and conclusions contained in this document are those of the author(s) and should not be interpreted as representing the official policies of DARPA or the U.S. government.

## CONTRIBUTIONS

D.A.S.H, K.L, R.C., and D. J. B. prepared the manuscript. D.A.S.H and K.L. designed the integrated photonic chips and K.D.N. fabricated them. D.A.S.H. built the ECTL, characterized the device, and performed all laser locking and frequency noise experiments associated with it. R.C. helped characterize and implement the modulation-free laser stabilization. K.L. built and packaged the SBS laser, characterized the device, and performed all laser locking and frequency noise experiments associated with it. All authors contributed to analyzing the experimental results. D. J. B. supervised and led the scientific collaboration.

## COMPETING INTERESTS

Dr. Blumenthal has consulted for Infleqtion and owns stock in the company. D. Heim, K. Liu, R. Chawlani, and K. Nelson declare no potential conflict of interest.

## SUPPLEMENTARY INFORMATION

**1. Introduction**

This supplementary information provides additional details on the design, performance, and operation of the stabilized stimulated Brillouin scattering (SBS) and stabilized external cavity tunable laser (ECTL) devices. This demonstration of circulator-free and isolator-free SBS is characterized in Supplementary Figs. 2, 3 & 4, and the wide tuning range of the ECTL, while maintaining narrow linewidth, is shown in Supplementary Fig. 5 & 6. Additional information regarding the modulation-free laser stabilization of the lasers to the 4-meter coil-loaded Mach Zehnder Interferometer (MZI) circuit is provided in Supplementary Sections 4, 5 & 6.

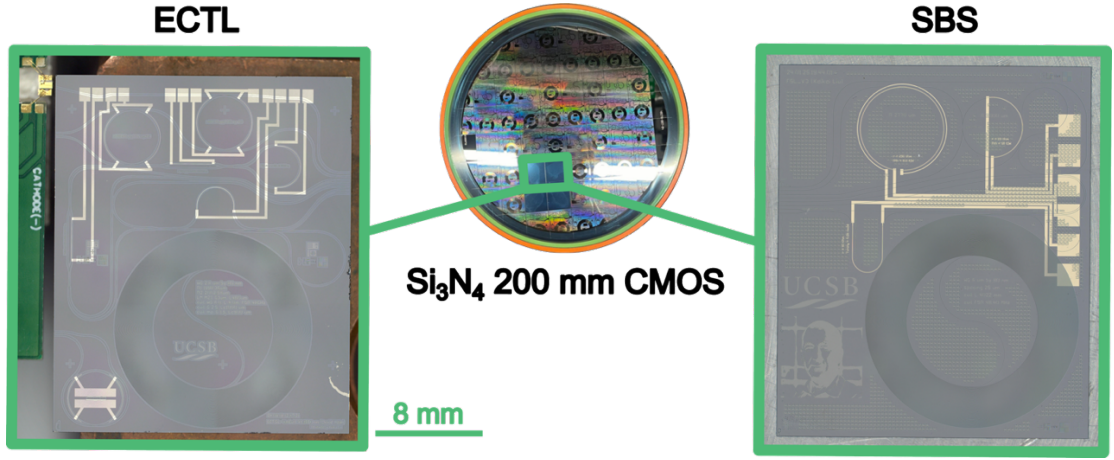

**Supplemental Fig. 1. Device images.** The stabilized external cavity tunable laser (ECTL, left) and stabilized stimulated Brillouin scattering laser (SBS, right) are both fabricated in an 80 nm thick silicon nitride ($Si_3N_4$) CMOS-compatible platform. This platform enables the design of low loss waveguides, high quality factor (Q) resonators, large mode volume reference cavities, and narrow linewidth hybrid-integrated lasers.

## 2. Circulator-free and isolator-free SBS laser design

The quality factor (Q) of the SBS resonator is a critical parameter that, combined with the mode volume, determines the lasing threshold of the device. Plotted in Supplementary Fig. 2a, we measure the SBS resonator with a loaded 16.2 million and intrinsic 39.0 million Q at 1550 nm with a total linewidth of 12.0 MHz, corresponding to a waveguide propagation loss of 0.8 dB/m. This enables a 6 mW lasing threshold and at a pump power of 25 mW the SBS laser outputs 9 mW at the first-order Stokes (S1) clamping point[1], shown in Supplementary Fig. 2b.

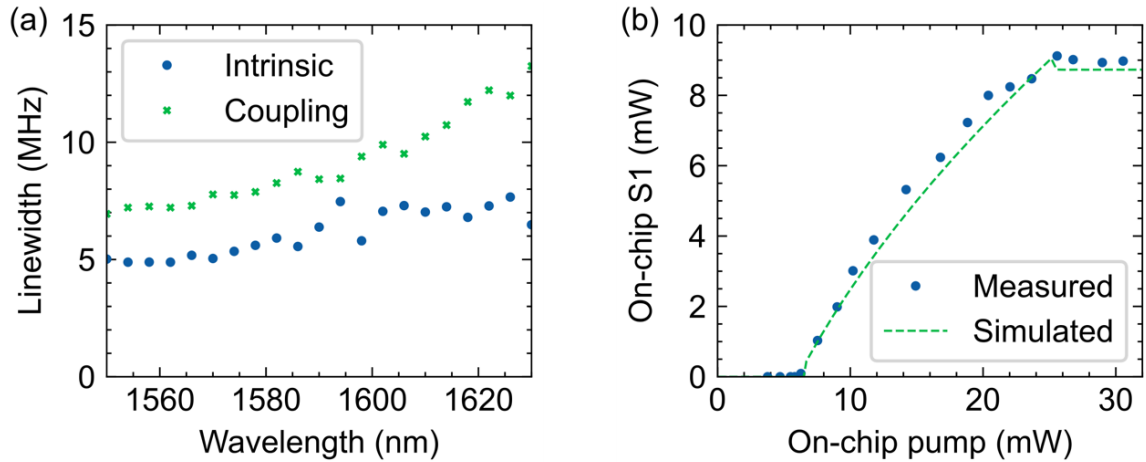

**Supplemental Fig. 2. Circulator-free SBS laser. a** Spectral scanning of the SBS resonator and add-through filter resonator while the filter resonator

The add-drop filter ring resonator, meanwhile, has a total linewidth of 0.7 GHz, as shown in Fig. 3a. Both the filter and SBS ring resonators have a micro-heater on top of them to tune the optical resonances thermo-optically. To operate the laser, the filter ring is tuned to align with the SBS resonance where the Stokes (S1) emission is emitted. Supplementary Fig. 3b compares the SBS laser output extracted by the on-chip filter with that obtained using a fiber-optic circulator[2], demonstrating a 23 dB suppression of unwanted back-reflected pump light.

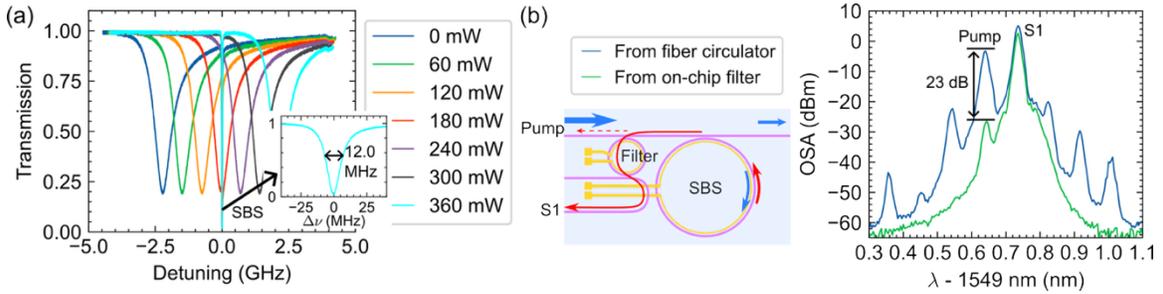

**Supplemental Fig. 3. Circulator-free SBS laser. a** Spectral scanning of the SBS resonator and add-through filter resonator while the filter resonator is being thermally tuned. **b** The SBS laser output dropped by the on-chip filter is compared to that dropped by a fiber-optic circulator without using the on-chip filter, showing a 23 dB suppression on the back-reflected pump power.

To test the SBS laser's resilience to external feedback, we experimentally inject laser light back into the output port of the laser and measure the frequency noise under different optical feedback levels. In the experiment outlined in Supplementary Fig. 4, 90% of the SBS laser is tapped and circulates back into the SBS laser using a fiber-optic circulator with a variable optic attenuator (VOA) that controls the feedback strength, which is defined as the ratio between the feedback power and the total SBS output power. The remaining 10% of the SBS laser output is used to monitor the laser frequency noise (FN) measured by an optical frequency discriminator (OFD) which is a fiber MZI with a free spectral range (FSR) of 1.027 MHz. With the feedback strength increasing from -43 dB to -7 dB, the frequency noise above 1 MHz is barely affected, and the fundamental linewidth remains the same at 4.0 Hz, but the frequency noise below 1 MHz slightly increases. The maximum -7 dB feedback strength is due to the insertion loss in the experimental setup. This resilience to optical feedback enables isolator-free operation of the laser and combining of the SBS with additional downstream integrated components such as the coil-loaded MZI reference cavity.

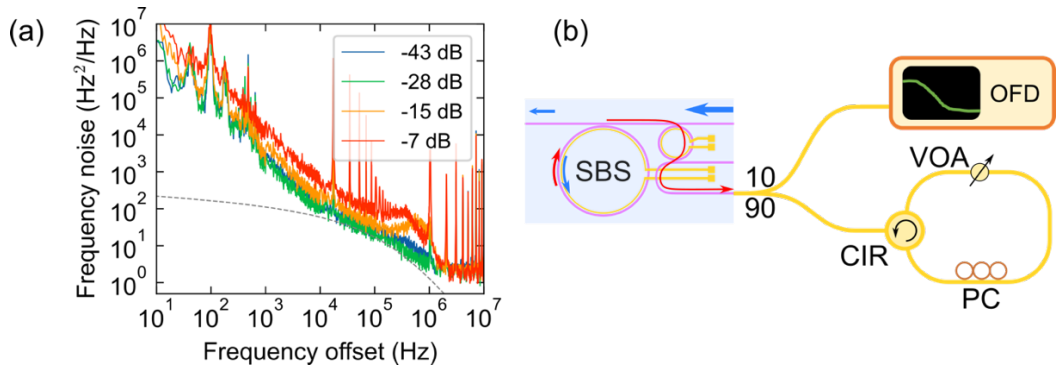

**Supplemental Fig. 4. SBS laser resilience to external feedback. a** Frequency noise of the SBS laser under varying optical feedback conditions, showing little to no effect on the fundamental linewidth of the laser due to feedback. **b** The optical feedback experiment where 90% of the SBS laser output circulates back into the SBS laser using a fiber-optic circulator (CIR) with a variable optic attenuator (VOA) and polarization control (PC) to adjust the feedback power level. The remaining 10% of the optical power is used to monitor the frequency noise measured by an optical frequency discriminator (OFD).

## 3. ECTL operation and frequency noise

The ECTL consists of an InP reflective semiconductor optical amplifier (RSOA) edge-coupled to a silicon nitride extended cavity photonic integrated circuit (PIC). The gain chip is coated with 90% reflectivity on the side opposite the PIC and an antireflection material on the near side with a reflectivity of 0.005%. It is then mounted on a temperature-controlled copper block for heat-sinking. The gain chip is wire bonded to a PCB that screws onto the copper block for external electrical control of the gain chip. The RSOA has an angled facet of 5.6° that requires the $Si_3N_4$ waveguide to be angled by 13.1° to best match the beam propagation direction. The $Si_3N_4$ PIC sits atop its own temperature-controlled mount and the RSOA is edge-coupled to the ECTL input waveguide which is 18 $\mu m$ wide at the facet and then tapers down to 2.8 $\mu m$ for the rest of the external cavity circuit. Thermo-optic metal heaters are deposited on the two intracavity Vernier rings for coarse laser tuning, on the Sagnac loop mirror MZI to adjust the reflectivity of the back mirror, and on a phase section for fine laser tuning. The ECTL operating wavelength as a function of current applied to one of the Vernier rings is plotted in Supplementary Fig. 5a. The average tuning efficiency of the Vernier tuning is measured to be 65 nm/W. We measure side mode suppression ratios (SMSRs) of up to 65 dB (Supplementary Fig. 5b) and fiber-coupled off-chip output power of 1 mW measured at 1540 nm.

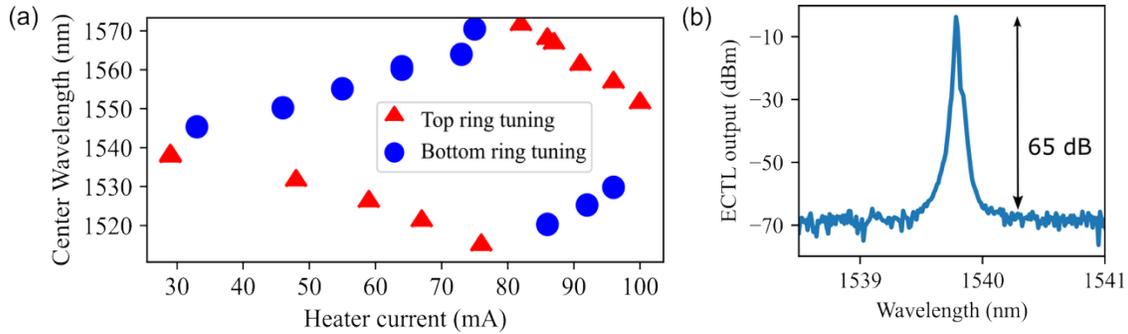

**Supplemental Fig. 5. ECTL single mode operation and tuning. a** The ECTL operating wavelength as a function of current applied to the microheater on one of the two intracavity Vernier rings. The laser tunes across ~ 60 nm with an average tuning efficiency of 65 nm/W. **b** The optical spectrum of the ECTL operating near 1540 nm as measured on an optical spectrum analyzer (OSA), showing an OSA noise-floor limited side mode suppression ratio (SMSR) of ~ 65 dB.

We measure the frequency noise of the freerunning ECTL across the tuning range using a 1.027 MHz unbalanced fiber MZI as an OFD. The frequency noise spectra are plotted in Supplementary Fig. 6a with measured fundamental linewidths of between 1.65 and 10.48 Hz measured across a wavelength range of 1515 – 1570 nm, and freerunning $1/\pi$ integral linewidths of between 935 and 2540 Hz. Weaker ring-bus coupling of the intracavity rings may contribute to longer effective cavity lengths and therefore a reduction in fundamental linewidth at shorter wavelengths.

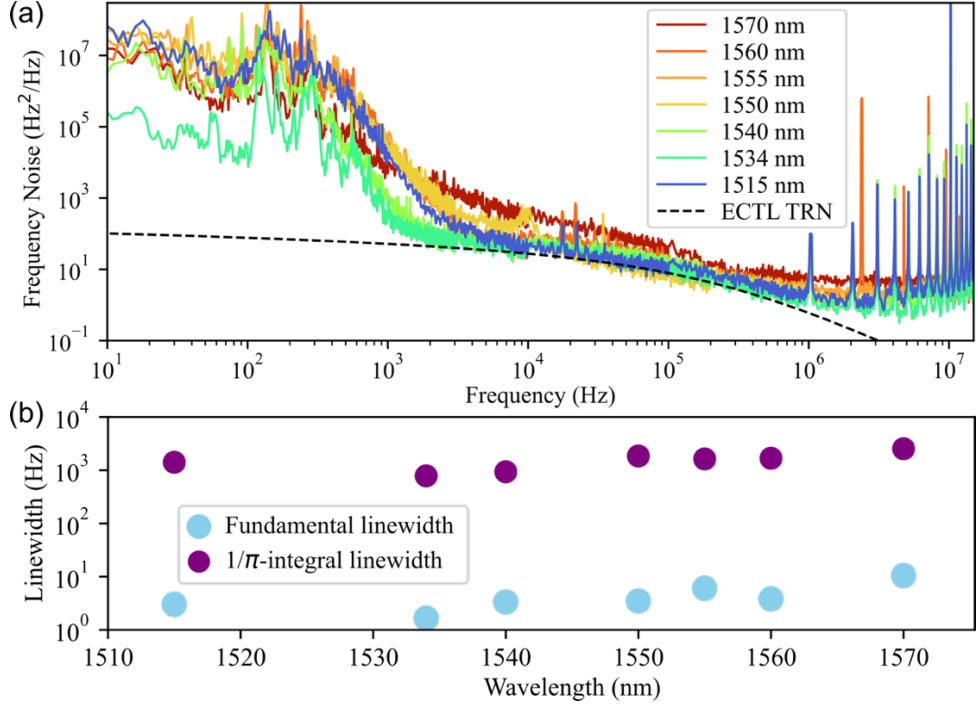

**Supplemental Fig. 6. Frequency noise of freerunning ECTL. a** Frequency noise of the freerunning ECTL measured using a 1.027 MHz unbalanced fiber MZI as an optical frequency discriminator. The dashed black curve is a calculated estimate of the thermorefractive noise (TRN) floor of the intracavity rings. **b** The fundamental and 1/π integral linewidth of the freerunning ECTL measured across the tuning range.

### 4. Modulation-free laser stabilization

We employ a modulation-free laser stabilization technique based on a resonator-loaded MZI where a 4-meter coil resonator is coupled to one arm of the MZI, illustrated in Supplementary Fig. 7a. The resonator-loaded MZI can be modeled in a matrix formulation by the following expression[3],

$$\begin{pmatrix} b_1 \\ b_2 \end{pmatrix} = \begin{pmatrix} \tau & i\kappa \\ i\kappa & \tau \end{pmatrix} \begin{pmatrix} T(\Delta\omega) & 0 \\ 0 & e^{i\Delta\phi} \end{pmatrix} \begin{pmatrix} \tau & i\kappa \\ i\kappa & \tau \end{pmatrix} \begin{pmatrix} a_1 \\ a_2 \end{pmatrix}$$

$$T(\Delta\omega) = \frac{i\Delta\omega + (\gamma_{ex} - \gamma_{in})/2}{i\Delta\omega + (\gamma_{ex} + \gamma_{in})/2}$$

where $T(\Delta\omega)$ is the add-through bus-coupled waveguide resonator transfer function, $\gamma_{ex}$ is the coupling loss rate, $\gamma_{in}$ is the intrinsic loss rate, and $e^{i\Delta\phi}$ represents the phase delay of the unbalanced MZI. This formulation assumes that the two directional couplers in the MZI each have two inputs and outputs and have the same coupling coefficient $\kappa^2$. Balanced detection of the two outputs can be expressed as $\Delta y = |b_2|^2 - |b_1|^2$ where for a phase delay of $\Delta\phi = \pi/2$ the output $\Delta y$ yields an asymmetric PDH-error-signal-like curve with a zero-DC quadrature point and maximum frequency discrimination slope (Supplementary Fig. 7b). We measure an error signal slope of 0.45 MHz/V when measured with the ECTL at 1540 nm (Supplementary Fig. 7c) and slopes of between 0.13 – 0.5 MHz/V over the 1520 – 1570 nm wavelength range.

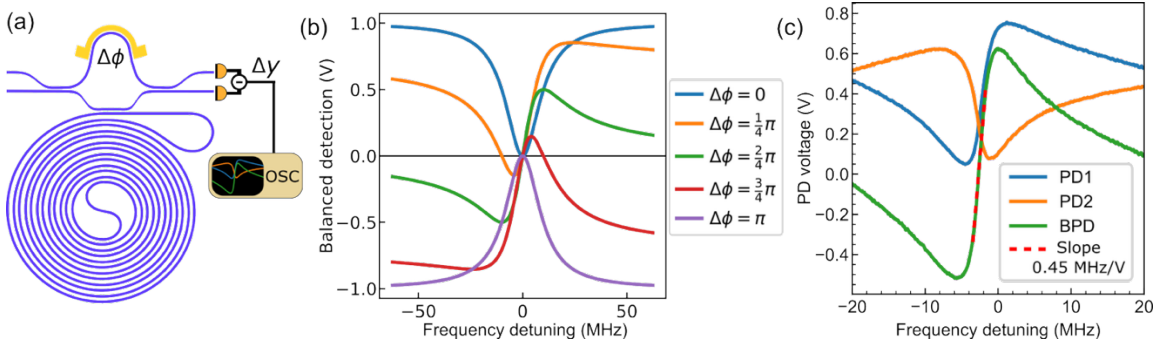

**Supplemental Fig. 7. Modulation-free last stabilization circuit. a** Schematic of the 4-meter coil-loaded Mach Zehnder Interferometer (MZI) with a tunable phase section and balanced detection of the two outputs generate the Pound-Drever-Hall (PDH) locking error signal. **b** Simulation of $\Delta y$ for various values of $\phi$. **c** Measurement data of the two coil-loaded MZI outputs (PD1 and PD2) and the corresponding balanced photodetector (BPD) signal from the 4-m coil-loaded MZI, probed with the external-cavity tunable laser (ECTL) near 1540 nm, demonstrating a frequency discrimination of 0.45 MHz/V with a zero-DC quadrature point.

### 5. ECTL laser stabilization

The ECTL utilizes current modulation of the RSOA as a high-speed frequency tuning mechanism to correct the laser frequency and we achieve modulation bandwidths of between 0.1 – 0.3 MHz. We measure the stabilized laser at three points across the nearly 60 nm tuning range and measure the frequency noise using two independent methods. The locked-laser frequency noise and the corresponding Allan deviation (ADEV) is shown in Supplementary Fig. 8a,b.

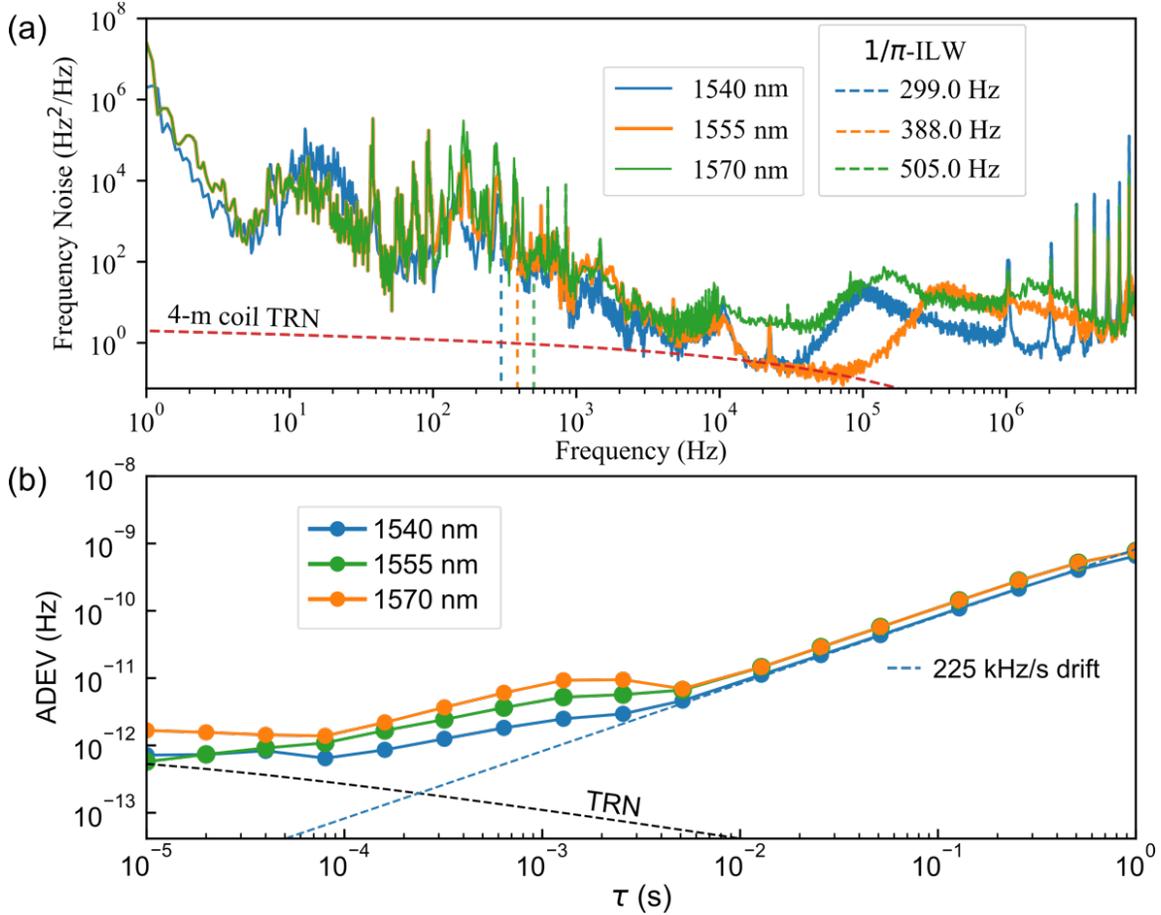

**Supplemental Fig. 8. Stabilized ECTL frequency noise. a** Frequency noise of the stabilized ECTL measured at three points across the tuning range. The results show $1/\pi$ integral linewidths (ILW), calculated using the reverse integration method, of 299, 388 and 505 Hz measured at 1540, 1555, and 1570 nm, respectively. **b** Allan deviation (ADEV) of the stabilized ECTL measured at three points across the tuning range.

### 6. SBS laser stabilization

To demonstrate laser stabilization of the SBS laser to the modulation-free 4-meter-coil resonator, we employ SBS laser pump power modulation as a fast feedback mechanism to steer the SBS laser optical frequency and the on-chip micro-heater on top of the SBS ring resonator as a slow and auxiliary feedback mechanism so that the pump power can remain relatively stationary and reduce the SBS laser intensity fluctuations. To test the speed of the pump-modulated SBS laser frequency response, an optical frequency discriminator (OFD) which is a 200-meter-delayline fiber MZI with an FSR of 1.03 MHz is used to measure the SBS laser frequency modulation response, and a vector network analyzer (VNA) outputs a modulation signal to an intensity modulator on the pump power and inputs the OFD signal from the SBS laser frequency modulation (Supplementary Fig. 9). The SBS laser frequency modulation response from the pump intensity modulation measured by the VNA exhibits a 180°-phase-lag frequency of ~0.8 MHz, which sets the upper bound of the laser stabilization bandwidth.

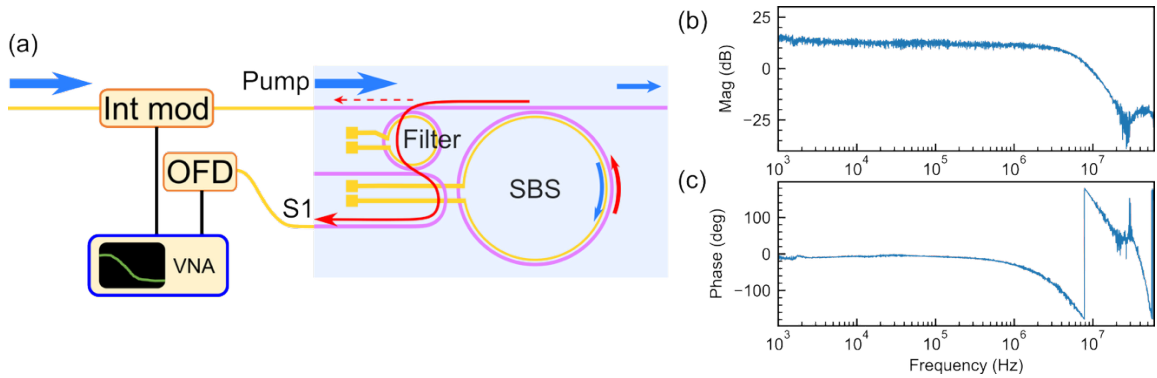

**Supplemental Fig. 9. Pump power frequency modulation of the SBS laser.** Intensity modulation of the pump laser is used as a fast feedback mechanism to steer the SBS laser optical frequency. **a** To test the speed of the pump-modulated SBS laser frequency response, a vector network analyzer (VNA) outputs a modulation signal to an intensity modulator on the SBS pump power, the SBS output (S1) is then input on a 1.03 MHz FSR fiber MZI, used as an optical frequency discriminator (OFD), to measure the SBS laser frequency modulation response on the VNA. **b** The frequency-modulation response to SBS pump-intensity modulation. **c** The phase-lag of the pump-intensity modulated SBS, exhibiting a 180° phase lag at ~0.8 MHz.